\documentclass[reqno,12pt]{amsart} 
\usepackage[top=1.5in,right=1.125in,left=1.125in,bottom=1.5in]{geometry}
\usepackage{amssymb}
\usepackage{color}
\usepackage{tikz}
\usepackage{soul}
\usetikzlibrary{positioning,decorations.pathmorphing}
\usetikzlibrary{positioning,decorations.pathmorphing}
\definecolor{red}{rgb}{0.7,0,0}
\definecolor{grey}{RGB}{112,112,112}
\definecolor{blue}{RGB}{034,113,179}
\usepackage[colorlinks=true,citecolor=blue,linkcolor=red,urlcolor=grey,backref=page]{hyperref}
\newcommand{\koniec}{\begin{flushright}  $\Box $ \end{flushright}}
\newtheorem{theo}{Theorem}[section] 
\newtheorem{prop}[theo]{Proposition}  
\newtheorem{lemma}[theo]{Lemma}

\theoremstyle{remark}

\def\theequation{\thesection.\arabic{equation}}

\newcounter{mnotecount}[section]

\renewcommand{\themnotecount}{\thesection.\arabic{mnotecount}}

\newcommand{\mnote}[1]
{\protect{\stepcounter{mnotecount}}$^{\mbox{\footnotesize
$
\bullet$\themnotecount}}$ \marginpar{
\raggedright\tiny\em
$\!\!\!\!\!\!\,\bullet$\themnotecount: #1} }

\newcommand{\HH}{\mathbb{H}}

\newcommand{\R}{\mathbb{R}}

\def\p{\partial}
\def\ov{\overline}
\def\be{\begin{equation}}

\def\ee{\end{equation}}

\def\bea{\begin{eqnarray}}
\def\eea{\end{eqnarray}}

\numberwithin{equation}{section}
\begin{document} \date{April 22nd, 2023}
\title{Elizabethan vortices}
\author{Maciej Dunajski}
\address{Department of Applied Mathematics and Theoretical Physics\\ 
University of Cambridge\\ Wilberforce Road, Cambridge CB3 0WA, UK.}
\email{m.dunajski@damtp.cam.ac.uk}
\author{Nora Gavrea}
\address{Clare College\\
University of Cambridge\\
Trinity Lane, Cambridge, CB2 1TL\\ UK.
}
\email{nag37@cam.ac.uk}
\maketitle
\begin{center}
{\em Dedicated to Nick Manton on the occasion of his 70th birthday.}
\end{center}

\begin{abstract}
Radial  solutions to the elliptic Sinh–Gordon and Tzitzeica equations can be interpreted as Abelian vortices on certain surfaces of revolution. These surfaces have a conical excess angle at infinity (in a way which makes them similar to Elizabethan ruff collars). While they can not be embedded in the Euclidean 3-space, we will show that they can be globally embedded in the hyperbolic space.
The existence of these hyperbolic embeddings follows from the asymptotic analysis of a Painlev\'e III ODE.
\end{abstract}
\section{Introduction}
The Abelian Higgs model at critical coupling admits static soliton solutions called vortices \cite{MS}. These vortices can be constructed
on any surface $\Sigma$ with a Riemannian metric $g$, and arise from solutions of the Taubes equation \cite{taubes_paper}
\be
\label{taubes_intro}
\Delta u=e^u-1.
\ee
Here $u$ is a function on $\Sigma$, and $\Delta$ is the Laplacian of $g$. This equation is valid outside the small discs enclosing the vortex positions,
where $u$ has logarithmic singularities. If $\Sigma$ is non--compact, then $u$ is required to tend to zero asymptotically, away from its singularities.

The Taubes equation is non--integrable on the plane $\Sigma=\R^2$ equipped with the flat metric. It was pointed out by Witten \cite{witten}, that (\ref{taubes_intro})
is integrable on the hyperbolic space with constant Gaussian curvature equal to $-1/2$, as then it can be transformed into the Liouville equation. There
are other vortex--type equations \cite{manton} (see also \cite{CD0, ross, sven}), where the RHS of (\ref{taubes_intro}) is replaced by $C_0-C_1e^u$ for some constants $(C_0, C_1)$. These equations
also reduce to the Liouville equation on spaces of constant curvature, which now depends on the combination of $(C_0, C_1)$. Like Witten's hyperbolic ansatz, the constant curvature
integrable cases
all originate from symmetry reductions of self--dual Yang--Mills equations on a four manifold $\Sigma\times S$, where $S$ is a surface of constant Gaussian curvature
equal to minus the Gaussian curvature of $\Sigma$ \cite{CD0}.

There are two more occurrences of integrability in the Taubes equation, which go beyond the Liouville equation, but instead transform 
(\ref{taubes_intro})  into the elliptic Sinh--Gordon equation, or the elliptic Tzitzeica equation \cite{D12}. To see how these equations arise, consider 
the metric $g$ in the complex isothermal coordinates, where $g=\Omega g_0$ with $g_0=dzd\ov{z}$, and allow the conformal factor $\Omega:\Sigma\rightarrow \R^{+}$ to depend on $u$. Choosing $\Omega=e^{-u/2}$ yields the Sinh--Gordon equation and $\Omega=e^{-2u/3}$ gives the Tzitzeica equation
\be
\label{SG_intro}
\Delta_0 \Big(\frac{u}{2}\Big)=\begin{cases}
               \sinh{\Big(\frac{u}{2}\Big)}\\
               e^{u/3}-e^{-2u/3}
            \end{cases}\quad\mbox{where}\quad \Delta_0=4\frac{\p^2}{\p z\p\ov{z}}.
\ee
If $u$ satisfies appropriate boundary conditions, then the corresponding vortex can be interpreted as a surface, with the Higgs field
and the  Abelian magnetic field encoded in the intrinsic Riemannian data of $g$.

If $u$ satisfies appropriate boundary conditions, then the corresponding vortex can be interpreted as a surface, with the Higgs field and the Abelian magnetic field encoded in the intrinsic
Riemannian data of $g$. This surface is however singular at the vortex position, which makes the
associated vortex number ill defined. We shall, for each positive integer $N$, construct another
surface $(\Sigma_N, g_N)$ (which for even $N$ is a ramified cover of $\Sigma$) which is regular and diffeomorphic
to $\R^2$.

The purpose of this paper is to visualise these vortices by constructing isometric embeddings of $(\Sigma, g)$. We shall show (see \S\ref{section_SG}) that for any non--negative
integer $N$, there exist solutions of the SG and Tzitzeica equation corresponding to circularly symmetric $N$--vortex solutions. The corresponding surfaces
can be embedded in $\R^3$ as surfaces of revolution with conical singularities at the position of the vortex. To resolve these conical singularities
we shall construct ramified covers of $\Sigma$, where the vortex metric is regular everywhere with the asymptotic behaviour
\[
g\sim\begin{cases}
                 dR^2+R^2d\psi^2 \quad\mbox{as}\quad R\rightarrow 0  \\
               dr^2+c_N^2 r^2d\psi^2\quad\mbox{as}\quad r\rightarrow \infty
					\end{cases}
						\]
where $r$ and $R$ are radial coordinates related by $r=R^{1+N/2}$ in the SG case, and $r=R^{1+2N/3}$ in the Tzitzeica case, and $c_N$ is a constant
equal to $1+N/2$ in the SG case, and $1+2N/3$ in the Tzitzeica case.
The metric is regular at the position $R=0$ of the vortex, and is locally asymptotically flat with the excess angle depending on $N$.
Surfaces with such asymptotic behaviour do not admit isometric embeddings in $\R^3$, at least as surfaces of revolution but it may be possible to construct global embeddings in $\R^3$ such that the extrinsic curvature only admits a discrete
isometry group. The resulting surfaces asymptotically look like the Elizabethan ruffs (Figure 1a) 
which can be locally unfolded into a flat region of a plane, but
contain too much material to do it globally: the circumference of circles exceeds $2\pi$ times the radius.
\begin{center}
{\includegraphics[width=6cm,height=5cm]{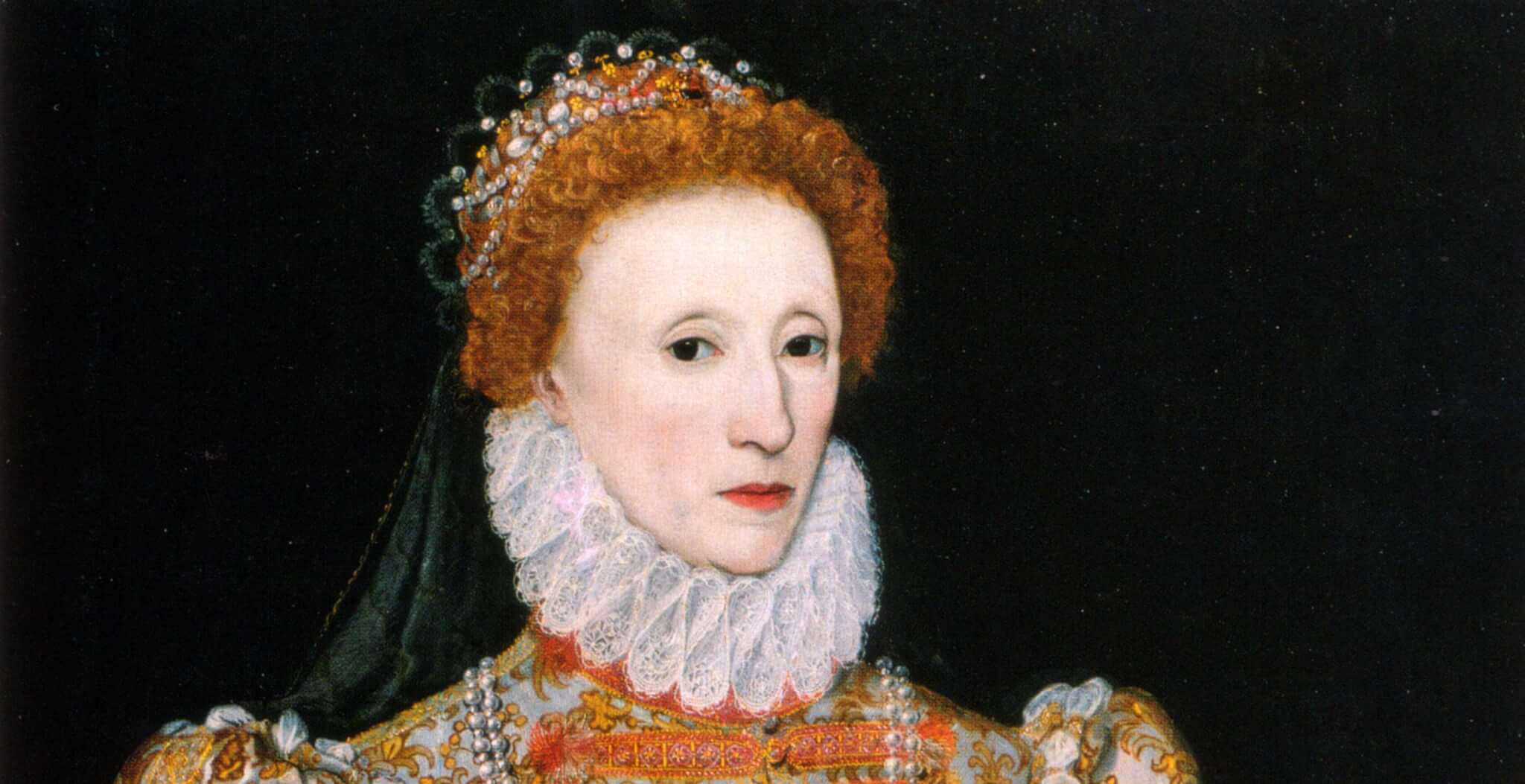}\;\; 
\includegraphics[width=6cm,height=5cm,angle=0]{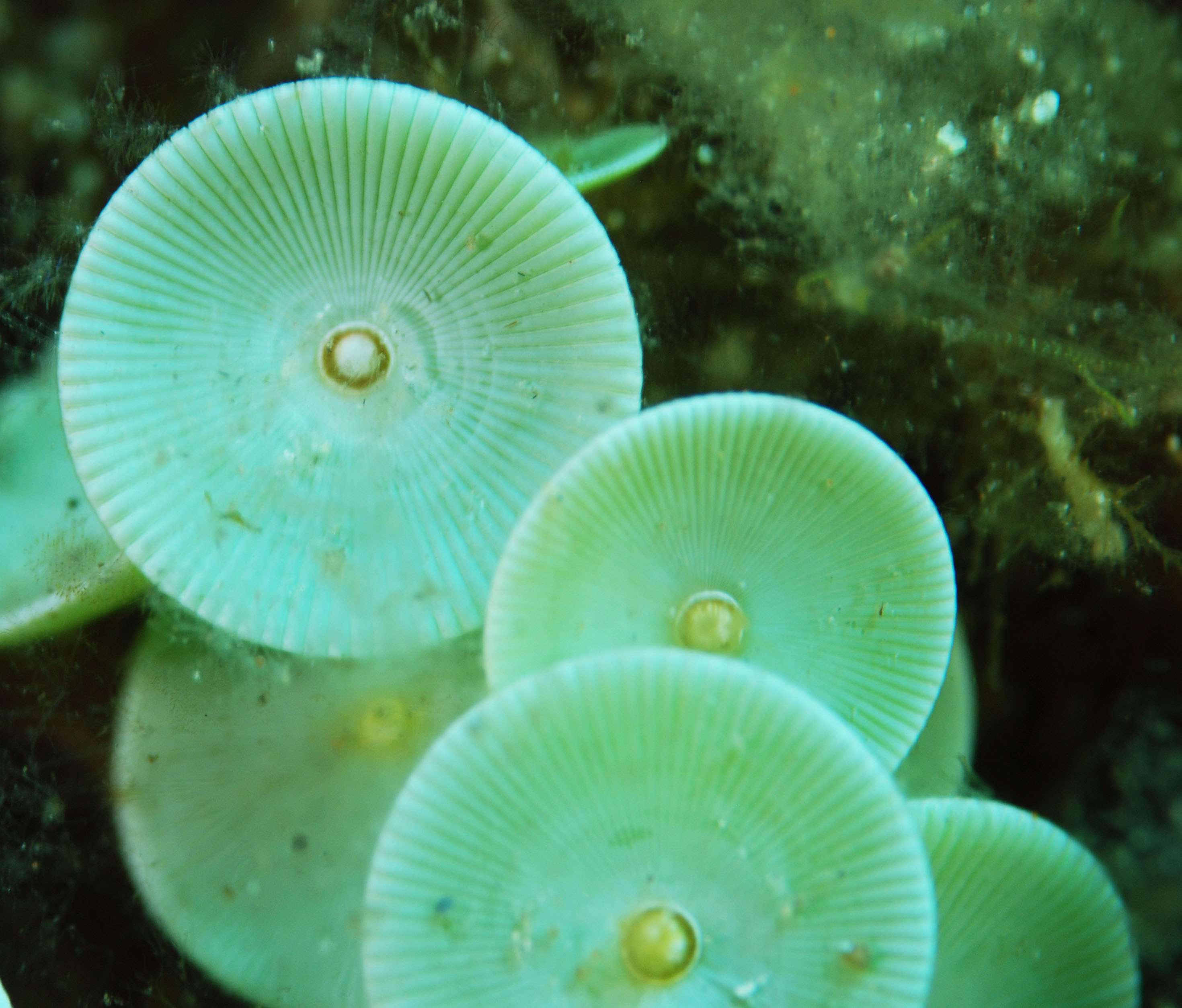}}
\end{center} 
{\em  {\bf Figure 1.} Asymptotic excess angle in the Elizabethan ruffs (1a), and green algae (1b).}
\vskip5pt
Motivated by this analogy we shall refer
to the surfaces resulting from (\ref{SG_intro}) as Elizabethan vortices. Other examples where the asymptotic circumference growth prevents the existence of
circularly symmetric embeddings are certain types of algae (Figure 1b).
The algae are in fact a better analogy to what happens with Elizabethan vortices, as
(unlike the hollow Elizabethan ruffs) their excess angle changes continuously with the radius. 
In \S\ref{section_he} we shall show that the Elizabethan vortices  embed as regular surfaces of revolution around a hyperbolic geodesic in the hyperbolic three--space. The existence and regularity of these embeddings are established in Theorems \ref{main_theorem} and \ref{main_theorem_tzitzeica} which are our main results. 

The mathematical theory of Abelian vortices and their moduli spaces has, over the last three decades, been advanced and developed by Nick Manton and his students and collaborators. It gives us a pleasure to dedicate this paper to Nick on the occasion of his seventieth birthday.
\subsection*{Acknowledgements} 
We are grateful to Rob Kusner, Nick Manton, and Marc Troyanov for useful discussions and correspondence.
We also thank the anonymous reviewers for their careful reading of the manuscript and many insightful suggestions. Nora Gavrea is grateful to have been supported by a CMS Bursary grant from the Faculty of Mathematics, University of Cambridge.
\section{The Taubes equation}
Let $(\Sigma, g)$ be a two--dimensional Riemannian manifold with the orientation given by the K\"ahler two--form $\omega_\Sigma$, 
and let ${\mathcal L}\rightarrow \Sigma$ be a Hermitian line bundle equipped with a $U(1)$ connection $A$, and a global
$C^{\infty}$ section $\phi$ called the Higgs field. 
We shall be interested in pairs $(A, \phi)$ which are global minimizers of the Ginsburg--Landau energy
functional at the critical coupling
\be
\label{DGfun}
E[A, \phi]=\frac{1}{2}\int_{\Sigma}\Big(|D\phi|^2+|F|^2+\frac{1}{4}(1-|\phi|^2)^2\Big) \omega_\Sigma
\ee
where $F=dA$, and $D\phi=d\phi -iA \phi$.
Completing the square in (\ref{DGfun}) shows that $E\geq \pi N$ where the integer
\[
N=\frac{1}{2\pi}\int_{\Sigma} F
\]
is the vortex number equal to the first Chern number of ${\mathcal L}$, and the inequality is saturated
iff $(A, \phi)$ satisfy the 1st order Bogomolny equations 
\[
\ov{\mathcal D} \phi=0, \quad F=\frac{1}{2}(1-|\phi|^2)\omega_{\Sigma}
\]
where $\ov{\mathcal D}$ is the anti--holomorphic part of the covariant derivative $D$. The Bogomolny equations, and the energy functional
(\ref{DGfun}) are invariant under gauge transformations
\[
\phi\rightarrow e^{i\alpha}\phi, \quad A\rightarrow A+d\alpha,
\]
where $\alpha:\Sigma\rightarrow\R$.

If $z=x+iy$ is a holomorphic isothermal coordinate, such that the metric on $\Sigma$ is
\[
g=\Omega(z, \ov{z}) dzd\ov{z}
\]
and, in a trivialisation of ${\mathcal L}$ the connection is given by $A=A_zdz+\ov{A_z}d\ov{z}$, then the first Bogomolny equation can be solved
to give $\ov{A_z}=-i\phi^{-1}\p_{\ov z}\phi$. Setting $\phi=e^{u/2+i\chi}$, where $u$ and $\chi$ are real functions on $\Sigma$, with $u$ being gauge--invariant,
reduces the second Bogomolny equation to a single 2nd order PDE called the Taubes equation \cite{taubes_paper}
\be
\label{taubes}
\Delta_0 u-\Omega(e^{u}-1)=0, \quad\mbox{where}\quad \Delta_0=4\p_z\p_{\ov{z}}.
\ee
This equation is valid outside small discs enclosing the logarithmic singularities of $u$ corresponding
to the positions of vortices where the Higgs field vanishes. If $|\phi|$ vanishes at $z_0$ with multiplicity $N_0$ then near $z_0$
the function $u$ has as an expansion of the form
\[
u=2N_0\ln{|z-z_0|}+\mbox{const}+\frac{1}{2}\ov{b}\cdot(z-z_0)+\frac{1}{2}b\cdot(\ov{z}-\ov{z_0})+\dots
\]
where the coefficients $b, \ov{b}$ depend on $z_0$ and $\ov{z_0}$.
The seminal result of Taubes is that the moduli space of solutions to (\ref{taubes}) with vortex number $N$ is a manifold of real dimension $2N$.
In other words, specifying the positions of vortices, and their multiplicities determines the solution uniquely.
\vskip5pt

Assume that the underlying surface admits a $U(1)$ isometry, so that $\Omega=\Omega(R)$ where $R^2\equiv |z|^2$.
If the solution of the Taubes equation only depends on $R$ in a way that $\phi$ vanishes at $R=0$ with multiplicity $N$, then (\ref{taubes})
reduces to an ODE \be
\label{taubes1}
\frac{d ^2u}{d R^2}+\frac{1}{R}\frac{d u}{d R}+\Omega(1-e^u)=0.
\ee
The recursion relations then give
\be
\label{smallR}
u=2N\ln{R}+b_0+b_1 R+ b_2 R^2+\dots\;, \quad \mbox{where}\quad b_1=0, \quad b_2=-\frac{\Omega(0)}{4}.
\ee
\section{The Sinh--Gordon vortex}
\label{section_SG}
 Taking $\Omega=\exp{(-u/2)}$ in the Taubes equation (\ref{taubes}) yields
the elliptic Sinh--Gordon equation
\be
\label{sinh}
\Delta_0\Big(\frac{u}{2}\Big)=\sinh\Big(\frac{u}{2}\Big).
\ee
This gives an interpretation of the metric $g$ as an isolated vortex. The
magnetic field and the Higgs field have an intrinsic geometric interpretation
as  the Riemann curvature two–form and the (inverse
of) conformal factor with a complex phase $\chi$, i.e.
\[
  g= e^{-u/2}dz d\overline{z}, \quad |\phi|^{2}=e^u,
  \quad A=\frac{i}{2}(\p-\overline{\p})u+d\chi
.\]
Choosing  a spin--frame ${\bf e}^1=e^{-u/4}dx, {\bf e}^2=e^{-u/4}dy$
such that $g=\delta_{ij}{\bf e}^i\otimes {\bf e}^j$,
the connection
and curvature forms of $g$ can be read off from the Cartan 
structure equations 
\[
d{\bf e}^i+{\Gamma^{i}}_j\wedge {\bf e}^j=0, \quad\mbox{and}\quad {R^i}_j=d{\Gamma^{i}}_j,
\]
and are given by
\begin{eqnarray*}
{\Gamma^{1}}_2&=&\frac{1}{4}(u_xdy-u_ydx)=-\frac{1}{2}A+\frac{1}{2}d\chi,\\ 
{R^1}_2&=&\frac{1}{4}(u_{xx}+u_{yy})dx\wedge dy=-\frac{1}{2}F. 
\end{eqnarray*}
To reinterpret the surface $(\Sigma, g)$ as a vortex we need
to construct a solution to (\ref{sinh}) which satisfies the boundary conditions
(\ref{smallR}). To do that we shall look for radial solutions of the form $u=u(r)$.
In this case the ODE
\be
\label{radial_SG}
u''+\frac{1}{r}u'=e^{u/2}-e^{-u/2}
\ee
resulting from (\ref{sinh}) is equivalent
to the Painlev\'e III equation  with parameters $(0, 0, 1, -1)$ 
(see Appendix). In \cite{D12} the Painlev\'e connection
formulae  of \cite{MTW77}  relating the asymptotic solution
to Painlev\'e III at $r=0$ and $r=\infty$ have been used to show that
\begin{eqnarray}\label{assymp_h}
u(r)&\sim&4\sigma \ln{r}+4\ln{\beta}-\frac{1}{4(1-\sigma)^2\beta^2}r^{2-2\sigma}\quad \mbox{for}\quad  r\rightarrow 0\\
&\sim& -8\kappa K_0(r)\quad \mbox{for}\quad r\rightarrow \infty\nonumber
\end{eqnarray}
with the connection formulae, valid for $0\leq \kappa \leq \pi^{-1}$
\be
\label{connection_formulae}
\sigma(\kappa)=\frac{2}{\pi}\arcsin{(\pi\kappa)},
\quad \beta(\kappa)=2^{-3\sigma}\frac{\Gamma((1-\sigma)/2)}{\Gamma((1+\sigma)/2)}
\ee
and so $0\leq\sigma\leq 1$. To construct an $N$--vortex solution on a regular surface for an arbitrary $N$ take\footnote{See \cite{CD} for another idea based on the Baptista superposition rule \cite{baptista}.}
\be
\label{sssigma}
\sigma=\frac{N}{N+2}.
\ee
Note that near $r=0$ the power series solution (\ref{assymp_h}) does not take the form (\ref{smallR}). This is because the conformal factor $\Omega=e^{-u/2}\sim r^{-2\sigma}$ is not regular at $r=0$. To resolve this singularity
define the new coordinate $R$ by
\be
\label{RR1}
r=R^{1+\frac{N}{2}}.
\ee
Now the asymptotic behaviour of $u$ and the metric near $R=0$ are 
\be
\label{uNv}
u=2N\ln{R}+4\ln{\beta}-\frac{(N+2)^2}{16\beta^2} R^2+\dots
\ee
and
\[
  g\sim B_N\Big(dR^2+\Big(\frac{2}{N+2}\Big)^2R^2 d\theta^2\Big),
\]
where $B_N$ is a constant dependent on $N$. This is in agreement
with (\ref{smallR}).
This metric has a conical singularity with the deficit angle $2\pi N/(N+2)$.
To obtain a regular surface we pass to a ramified covering surface $\Sigma_N$ of $\Sigma$ taking the period
of $\theta$ to be $(2+N)\pi$, so that 
\[
\psi=\frac{2\theta}{2+N}
\]
is periodic with period $2\pi$. If $N$ is even, then
$\Sigma_N$ is a ramified cover of $\Sigma$.
Near $R=0$ the metric on $\Sigma_N$ is regular 
but near $R=\infty$ it has an excess angle as then $u\rightarrow 0$ and up to exponentially small corrections 
\begin{eqnarray}
\label{g_infty1}
g_N&\sim& B_N(dR^2+R^2d\psi^2)\quad\mbox{as}\quad R\rightarrow 0\nonumber\\
g_N&\sim& dr^2+\Big(\frac{N+2}{2}\Big)^2 r^2d\psi^2\quad\mbox{as}\quad r\rightarrow \infty.
\end{eqnarray}
The metric $g_N$ is curved between $R=0$ and $R=\infty$ with the Gaussian curvature
$
  K=(e^u-1)/4
$
interpolating between
$-1/4$ and $0$.

The following Lemma uses the maximum principle to establish the monotonicity of $u$.
\begin{lemma} 
\label{lemmavh}
Let $u=u(r)$ satisfy the radial Sinh--Gordon equation (\ref{radial_SG}), {with the asymptotic behaviour given by (\ref{assymp_h}).}
Then $u'(r)>0$ for $r\in (0, \infty)$.
\end{lemma}
\noindent
{\bf Proof.} Near $r=0$ we have $u'\sim 4N/(r(N+2))>0$, so for the statement of the Lemma not to be true there must exist $r_0$ s.t. $u'(r_0)=0$. Let us assume that
$r_0$ is the smallest value of $r$ such that $u'(r)=0$. {If \(u''(r_0)=0\), then equation (\ref{radial_SG}) implies that \(u(r)=0\).}
If $u(r_0)>0$ 
then $u$ must admit a local maximum with
$u''(r_0)<0$ and we get a contradiction as the RHS of  (\ref{radial_SG}) is positive, but the LHS is negative.
If $u(r_0)<0$ then there must also exist an $r_1$ such that $u(r_1)$ is a local minimum (as otherwise the function could not tend to $0$). Therefore
$u''(r_1)>0$, and now the RHS  (\ref{radial_SG}) is negative
but the LHS is positive. If $u''(r_0)=0$ then we also reach a contradiction
regardless of the sign of $u(r_0)$ as long as $u(r_0)\neq 0$.
Finally if $u(r_0)=0$ and $u'(r_0)=0$ then the initial value problem for (\ref{radial_SG}) at $r_0$ has a solution
$u\equiv 0$ but we  know that our solution is non--constant.
\koniec
\begin{center}
{\includegraphics[width=10cm,height=6cm,angle=0]{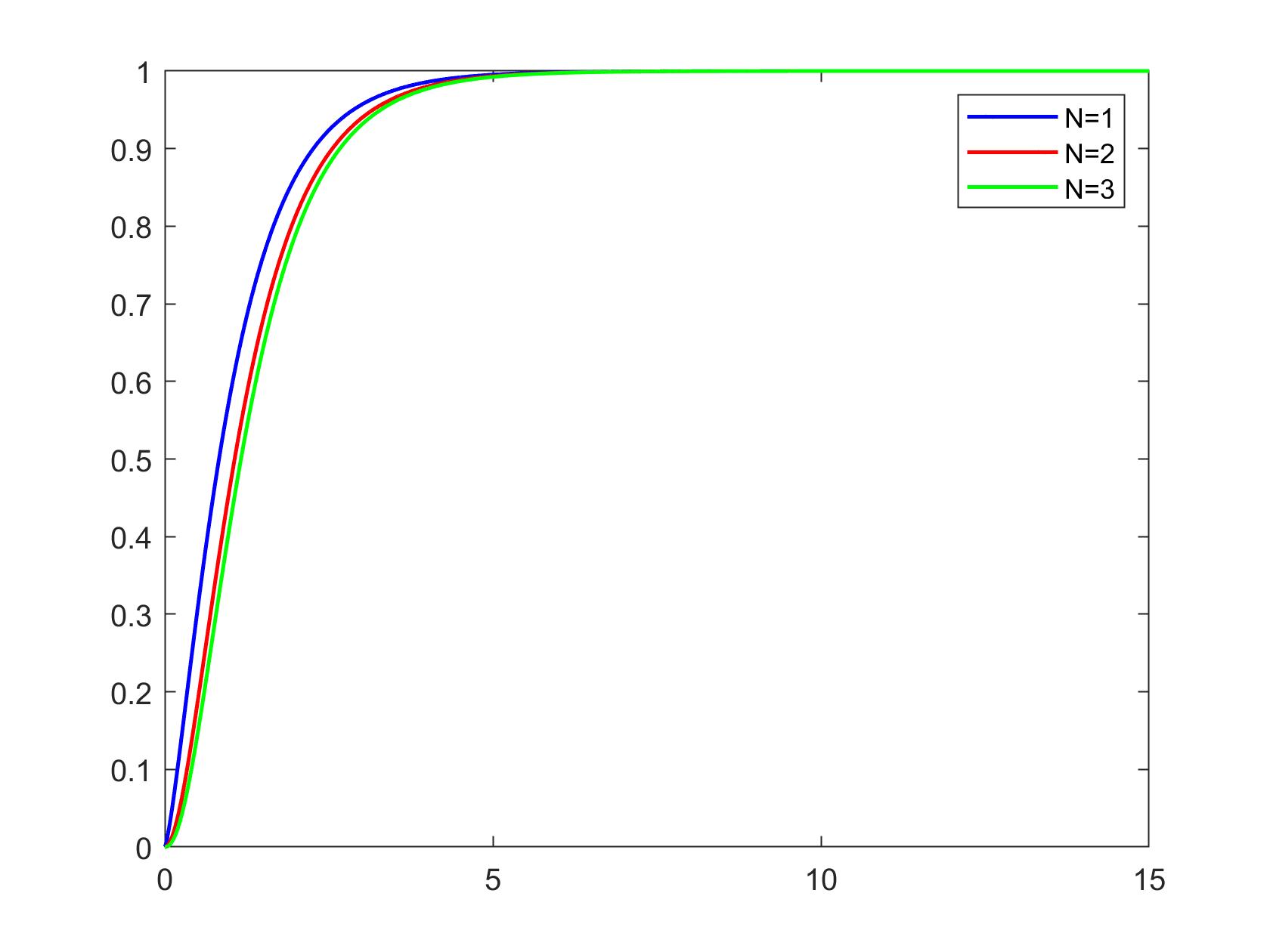}}
\end{center}
\nopagebreak
{\em  {\bf Figure 2.} Profile
 of $|\phi|^2=e^u$ for vortex numbers $N=1, 2, 3.$}

\vskip 5pt
The profiles of $u$ in Figure 2, and the profiles of the embeddings in Figures 3, 5, 6 were obtained numerically. When solving (\ref{taubes1}) subtract the logarithmic singularity
by setting $u(R)=h(R)+2N\ln{(R)}$, and shoot
on $b_0$ with the initial conditions $ h(0)=b_0, h'(0)=0$. In the case of the Sinh--Gordon and Tzitzeica vortices, the resulting ODE has a removable singularity at $r=0$. To get around this either change the coordinates to $R$ as in (\ref{RR1}) and (\ref{RR2}), or specify the initial conditions at $r=\epsilon$ (we took $\epsilon=10^{-30}$). We used both approaches, and applied
MATLAB ODE solver, ode45, which is based on a Runge-Kutta method with a variable time-step.
\subsection{Large $r$ asymptotics}
For large $r$ we approximate $\sinh(u/2)\sim u/2$, and deduce that $u$ is proportional
to a Bessel function of order $0$.
\be
\label{bess}
u''+\frac{1}{r}u'=u, \quad\mbox{so that}\quad u=cK_0(r)\sim c\sqrt{\frac{\pi}{2r}}e^{-r} \quad \mbox{as}\quad r\rightarrow \infty,
\ee
where $c$ is a constant.
In the particle interpretation of Speight \cite{speight_paper} a vortex asymptotically behaves as a point-like object carrying both a scalar charge, and a magnetic dipole moment. The strength of this point--like charge is given by the absolute value of the constant $c$. In the Sinh--Gordon case we can read--off the exact value of this 
constant from the connection formulae  (\ref{connection_formulae}) with $\sigma=N/(N+2)$ which yields
\[
|c|=\frac{8}{\pi}\sin{\Big(\frac{\pi N}{2(N+2)}\Big)}.
\]
For comparison, the strength of a $1$--vortex on the plane can only be computed numerically, and we found that
\[
|c|=\lim_{r\rightarrow\infty} \frac{u(r)}{K_0(r)}\sim 3.41.
\] 
In the original paper of Speight \cite{speight_paper} this numerical value was computed to be $3.36$.
\section{Sinh--Gordon isometric embeddings}
We turn to visualizing $\Sigma$ and $\Sigma_N$ as surfaces. The original surface $\Sigma$ can be embedded isometrically in $\R^3$
as a surface of revolution which is regular apart from the conical singularity at $R=0$. The ramified covering surfaces $\Sigma_N$
can also be embedded as  regular surfaces
of revolution in the hyperbolic space. We shall discuss these embeddings in
turn.
\subsection{Euclidean embedding}
\begin{prop}
  \label{prop_euclidean_SG}
There exists an isometric embedding $\iota: \Sigma\rightarrow \R^3$ of
the Sinh--Gordon $N$--vortex as a surface of revolution which is asymptotically
flat, and regular everywhere  away from the conical singularity {with deficit angle \(2\pi N/(N+2)\)}.
\end{prop}
\noindent
{\bf Proof.}
We shall prove this by constructing this embedding explicitly. Consider
the flat metric on $\R^3$ in the cylindrical coordinates $(\rho, \theta, z)$
and pull it back to $\Sigma$. Assuming that the embedding preserves the $U(1)$
symmetry, we must have $z=z(r)$ and $\rho=\rho(r)$ so that
\[
d\rho^2+\rho^2 d\theta^2+dz^2=[(\rho')^2+(z')^2]dr^2+\rho^2d\theta^2 =e^{-u/2}(dr^2+r^2d\theta^2)
\]
\be
\label{rhoz}
\rho=e^{-u/4}r, \quad z'=e^{-u/4}\sqrt{\frac{1}{2}ru'\Big(1-\frac{1}{8}ru'\Big)}.
\ee
For this to work we need the argument of the square root to be non--negative. As $u'>0$ (see Lemma \ref{lemmavh}), a problem can only arise if $u'>8/r$.
To rule this out note that (\ref{assymp_h}) and (\ref{sssigma}) imply
\[
\lim_{r\rightarrow 0} ru'=\frac{4N}{N+2}, \quad\mbox{and}\quad \lim_{r\rightarrow \infty} ru'=0.
\]
To compute the maximum of $ru'$ we look at its critical points
\[
0=(ru')'=r(u''+u'/r)=r(e^{u/2}-e^{-u/2})
\]
so $r=0$ or $u=0$ (so $r=\infty$). We conclude that  $ru'\leq 4N/(N+2)<8$. Therefore a global embedding in $\R^3$ exists away from the conical singularity at $r=0$.
\koniec
Expressing the conformal factor near $r=0$ in terms of $R$ yields
\be
\label{near0}
g\sim\Big(\frac{(N+2)^2}{4\beta^2}+\frac{(N+2)^4R^2}{128\beta^4}\Big)\Big(dR^2+
\Big(\frac{2}{N+2}\Big)^2R^2d\theta^2\Big), \quad\mbox{where}\quad
r=R^{1+\frac{N}{2}}.
\ee
The embedding equations (\ref{rhoz}) can be solved for small $R$, which yields
\begin{eqnarray*}
\rho&=&\frac{R}{\beta}+\frac{(N+2)^2}{64\beta^3}R^3+\dots,\\ 
z&=&\frac{\sqrt{N(N+4)}}{2\beta}R+\frac{\sqrt{N(N+4)}(N+2)^2(N^2+4N-8)}{384\beta^3(N+4)N}R^3+\dots
\end{eqnarray*}
in agreement with (\ref{near0}). 
Figure 3 contains plots of the embedded surfaces for $N=1, 2, 3$, {obtained numerically}.
\begin{center}
{\includegraphics[width=10cm,height=8cm,angle=0]{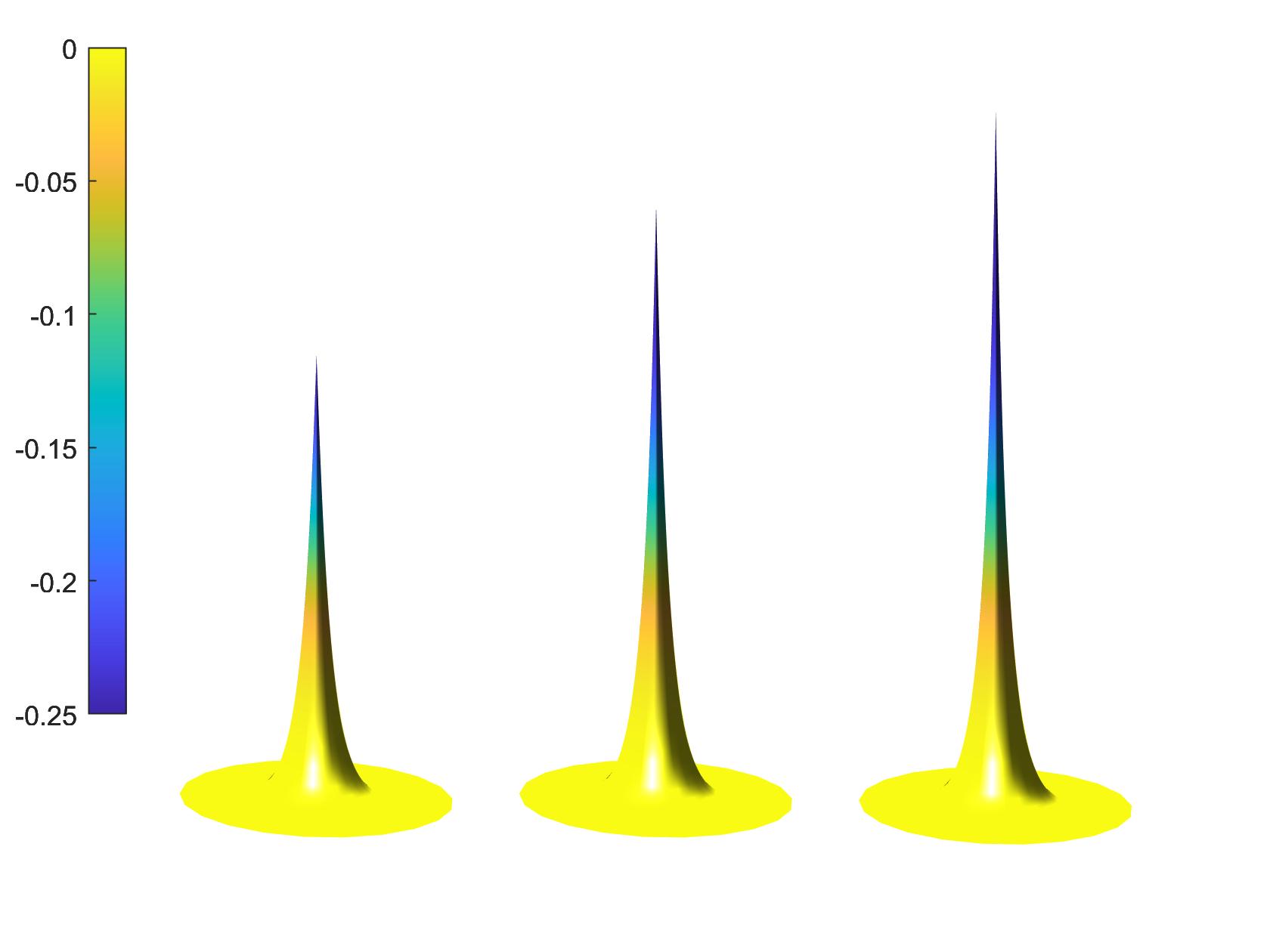}}
\end{center}
{\em  {\bf Figure 3.} Isometric embeddings of $\Sigma$ in $\R^3$ for $N=1, 2, 3$ coloured by the Gaussian curvature.}
\vskip5pt
The conical singularity of the embedding is reflected in the blow-up of the mean curvature $H$ at $r=0$. {The Gaussian curvature $K$} is everywhere regular (Figure 4).
\[
H=\frac{e^{3u/2}}{2r^2}(\rho z'+ \rho'z''-\rho''z'), \quad K=\frac{1}{4}(e^u-1).
\]
\begin{center}
  {\includegraphics[width=10cm,height=8cm,angle=0]{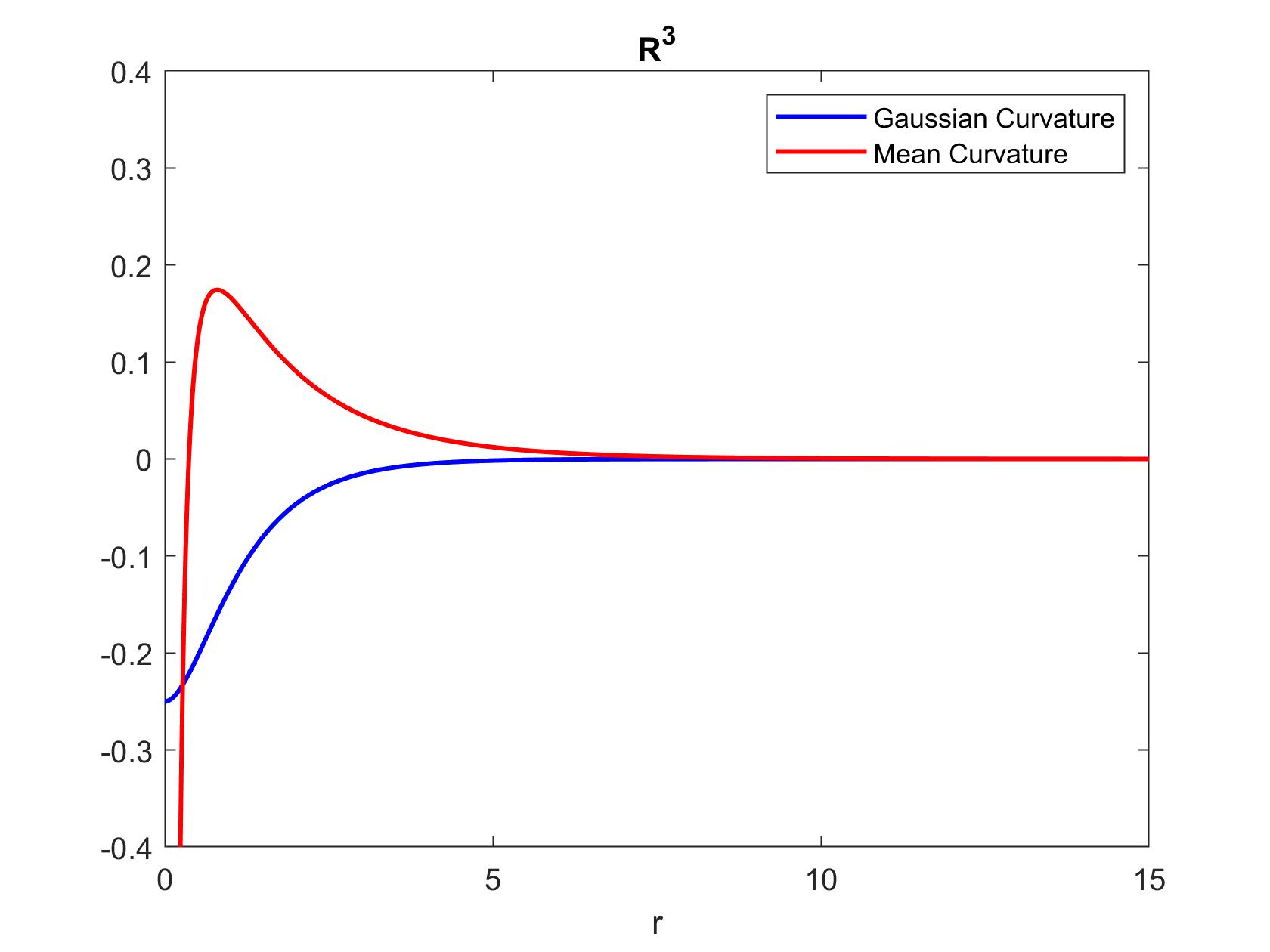}}
  \end{center}
  {{\em  {\bf Figure 4.} Mean and Gaussian curvatures of the isometric embeddings in $\R^3$ for $N=2$.}}
  \subsection{Hyperbolic embeddings}
  \label{section_he}
The regular ramified surface $\Sigma_N$ can not be embedded in $\R^3$ as a surface of revolution.  The circumference of circles 
centered at $R=0$ in $\Sigma_N$ grows faster than their radii, which makes such an embedding impossible. To see it explicitly, consider the asymptotic form
(\ref{g_infty1}) of the metric on $\Sigma_N$. Its pull-back to $\R^3$ with cylindrical
polars yields $\rho=r(N+2)/2$, so that
\[
  (z')^2=1-\Big(\frac{N+2}{2}\Big)^2
\]
which does not have solutions if $z$ is real, and $N>0$.

These surfaces can however, for any $N$, be properly embedded in the hyperbolic space $\HH^3$ as surfaces of revolution around a hyperbolic geodesic. Before establishing this fact
(Theorem \ref{main_theorem}) we shall demonstrate
that a local embedding of a surface of revolution in the hyperbolic space is unique, up to a hyperbolic isometry. In what follows, we shall use two models of the hyperbolic 3--space with scalar curvature $-6/L^2$: the upper half-space model with the metric
\be
\label{H3met1}
G_{\HH^3}=L^2\frac{dz^2+d\rho^2+\rho^2 d\psi^2}{z^2}, \quad z>0
\ee
and the Poincar\'e ball model with the metric
\be
\label{H3met2}
G_{B}=4L^2\frac{dw^2+w^2(d\chi^2+{\sin^2{\chi}}d\psi^2)}{(1-w^2)^2}.
\ee
These models are related by 
\be
\label{kusner}
z=\frac{1-w^2}{1+w^2-2w\cos\chi} ,\quad \rho=\frac{2w\sin{\chi}}{1+w^2-2w\cos{\chi}}.
\ee
\begin{prop}
  \label{prop_2in1}
A Riemannian surface with $U(1)$ isometry given in the arc-length parametrisation as 
\be
\label{revolution}
  g=dr^2+f(r)^2d\psi^2.
\ee
admits a local embedding as a surface of revolution in the hyperbolic space $\HH^3$ around a hyperbolic geodesic $\gamma$ in the range of $r$ given by
\be
\label{H3const}
L^2(1-(f')^2)+f^2\geq 0.
\ee
Given $\gamma$,  this embedding is unique up to a discrete hyperbolic isometry.
\end{prop}
\noindent
{\bf Proof.}
Considering the pull back of the hyperbolic metric (\ref{H3met1}) gives
two hyperbolic embeddings in $\HH^3$
\be
\label{two_emb}
  z_{\pm}(r)= \exp{\Big(\int_0^r \frac{-ff'\pm\sqrt{L^2(1-(f')^2)+f^2}}{L^2+f^2} du\Big)}, \quad \rho_{\pm}(r)=\frac{z_{\pm}(r) f(r)}{L},
\ee
which are well defined as long as (\ref{H3const}) holds.

To show that the two embeddings $(z_{\pm}(r), \rho_{\pm}(r))$ are related by a discrete
hyperbolic isometry, consider instead the pull back of the Poincar\'e ball metric (\ref{H3met2})
which gives
  \[
    f= 2L\frac{w\sin{\chi}}{1-w^2}, \quad 1=4L^2\frac{(w')^2+w^2{\chi'}^2}{(1-w^2)^2}.
  \]
  Set $Z=we^{i\chi}$, then
  \[
    f=L\frac{Z-\bar{Z}}{i(1-|Z|^2)}, \quad 1=\frac{4L^2|Z'|^2}{(1-|Z|^2)^2}
  \]
  so that if $Z$ is a solution, then so is $-\bar{Z}$. This corresponds to $\chi$ and $\pi-\chi$.
\koniec
As an example, consider $f=kr  $ which is the $k$--fold covering of the plane. The
hyperbolic embeddings (\ref{two_emb}) exist as long as
\[
r^2>\frac{L^2(k^2-1)}{k^2}
\]
which can be made as small as we want by the choice of $L^2$, but will not cover
$r=0$. Thus the $k$-fold cover of a plane can not be embedded in $\HH^3$ globally.
On the other hand a surface interpolating between a plane, and its $k$th cover with a profile function
\[
  f=r\sqrt{\frac{1+k^2r^2}{1+r^2}}
\]
admits a global embedding in $\HH^3$.
\begin{theo}
  \label{main_theorem}
Let $(\HH^3, G)$ be a hyperbolic space with the metric
  $G$ with Ricci scalar $-6/L^2$. For any $L^2\in (0, 4]$
  there exists
  a regular isometric embedding $\iota:{\Sigma_N}\rightarrow\HH^3$
  which preserves the $U(1)$ symmetry of $g_N$, and is unique up to a hyperbolic isometry.
\end{theo}
\noindent
{\bf Proof.}
We shall consider ${\Sigma_N}$ as a surface of revolution
around a geodesic $\gamma$ in $\HH^3$. We shall work in the upper half--space
model $z>0$  with the metric $G_{\HH^3}$ given by (\ref{H3met1})
and choose $\gamma$ to be the $z$ semi--axis as in the proof of Proposition \ref{prop_2in1} . The pull back of $G$ to $\Sigma_N$
\[
  \iota^*(G)=e^{-u/2} \Big(dr^2+\Big(\frac{N+2}{2}\Big)^2 r^2d\psi^2\Big)
\]
yields the embedding formulae 
\begin{eqnarray}
\label{z_emb}
  z_\pm&=&z_0\exp{\Big(\int_0^r\frac{(N+2)^2(t^2u'-4t)e^{-u/2}
       \pm\sqrt{e^{-u/2} P}}{4(N+2)^2t^2e^{-u/2}+16L^2}dt \Big)},\nonumber\\
\rho_\pm&=&\frac{N+2}{2L}rz_{\pm}e^{-u/4},
\end{eqnarray}
where
\[
P\equiv {16t^2e^{-u/2}(N+2)^2-L^2((N+2)tu'-4N)((N+2)tu'-4N-16)}.
\]
For this embedding to be regular we need $P\geq 0$. First compute the asymptotic behaviour
\begin{eqnarray*}
P&\sim& 16((N+2)^2t^2-N(N+4)L^2)\rightarrow \infty\quad \mbox{as}\quad t\rightarrow\infty\\
P&\rightarrow& 0\quad\mbox{as}\quad r\rightarrow 0.
\end{eqnarray*}
Next look for the critical points of $P$
to show that it does not have any and therefore stays positive between
$0$ and $\infty$. Using (\ref{radial_SG}) we find
\[
P'=2r(N+2)^2(L^2e^{u/2}+(4-L^2)e^{-u/2})(4-ru')  
\]
which clearly vanishes at $r=0$ and is non--negative if $L^2\leq 4$. Moreover, if $L^2=4$, then we avoid a blow up of $P'$ at $r=0$ if $N>2$. 
We claim that $Q\equiv 4-ru'>0$. Indeed, $Q(0)=8/(N+2)$
and $\lim_{r\rightarrow\infty} Q=4$ and 
  \[
    {Q}'=-r(e^{u/2}-e^{-u/2})
  \]
  which is zero only if $r=0$ or $u=0$. But in the proof of Lemma \ref{lemmavh}
  we have shown that $u\neq 0$ for finite $r$.
The two signs
  in (\ref{z_emb}) correspond to two embeddings related by a hyperbolic isometry as shown in Proposition \ref{prop_2in1}.
\koniec
{\bf Remarks}
  \begin{itemize}
\item The map $\iota:\Sigma_N\rightarrow \HH^3$ defined in Theorem
  \ref{main_theorem}
  is a proper embedding (rather than an immersion); there are no self--intersections in the hyperbolic space which can intuitively
be explained as the circumference of the hyperbolic circle grows faster than $2\pi$ times the hyperbolic radius in a way
which is sufficient to accommodate the $k$--fold  cover
of $\R^2$ asymptotically.

\item
Near the vortex position $r=0$ the embedding (\ref{z_emb}) is
\be
\label{zero_asy}
z\sim 1-\frac{(N+2)^2}{32\beta^2}r^{4/(N+2)}, \quad \rho\sim \frac{N+2}{4\beta}r^{2/(N+2)}, \quad \mbox{so that}\quad z\sim 1-\frac{\rho^2}{2},
\ee
where we used a hyperbolic isometry (scaling)  to set $z(0)=1$. 
{The particular choice $L^2=4$ is noteworthy in the sense that in this case} embedding osculates $\HH^2\subset \HH^3$ which embeds in the hyperbolic space as the hemisphere
$z^2+\rho^2=1$. Indeed, near $z=1$ this hemisphere is
\[
z=\sqrt{1-\rho^2}\sim 1-\frac{\rho^2}{2}
\]
in agreement with (\ref{zero_asy}). In particular this reaffirms the regularity
of the embedding at $r=0$ or $\rho=0$, as at this point the revolving curve
$z=z(\rho)$ has vanishing gradient.
\item
Near $r=\infty$ (choosing one of the two embeddings) both $z$ and $\rho$ tend to 
an annular limit point  $(z=0, \rho=0)$ where the surface meets the boundary of $\HH^3$ tangentially with a rate
\[
z\sim r^{-\frac{N+4}{N+2}}, \quad \rho\sim \frac{N+2}{4}r^{-\frac{2}{N+2}}, \quad\mbox{so that} \quad 
z\sim  \frac{(N+2)^{\frac{N+4}{2}}}{2^{N+4}} \rho^{\frac{N+4}{2}}.
\]
This asymptotes a cover of a horosphere\footnote{Recall that in the upper half space model a horosphere is either a sphere tangent to the boundary, or a plane parallel to the boundary. The horospheres have constant, non--zero, mean curvature  $H$ (CMC) in $\HH^3$ but are intrinsically flat with zero Gaussian curvature.
On the other hand 
the Euclidean planes and hemispheres in $\HH^3$ which intersect
the boundary orthogonally are isometric to $\HH^2$, and have non--zero constant Gaussian curvature \(K\), but vanishing mean curvature $H$.
The hemispheres which intersect the boundary transversally have $0<H<1$.} in $\HH^3$ as $r\rightarrow \infty$, with the metric (\ref{g_infty1}). On 
Figures 5  and 6 we present the surface corresponding to $N=2$ in the upper half space as well as
the Poincar\'e ball model of the hyperbolic geometry, together
with the plane sections.
\begin{center}
  \includegraphics[width=5.5cm,height=4.5cm,angle=0]{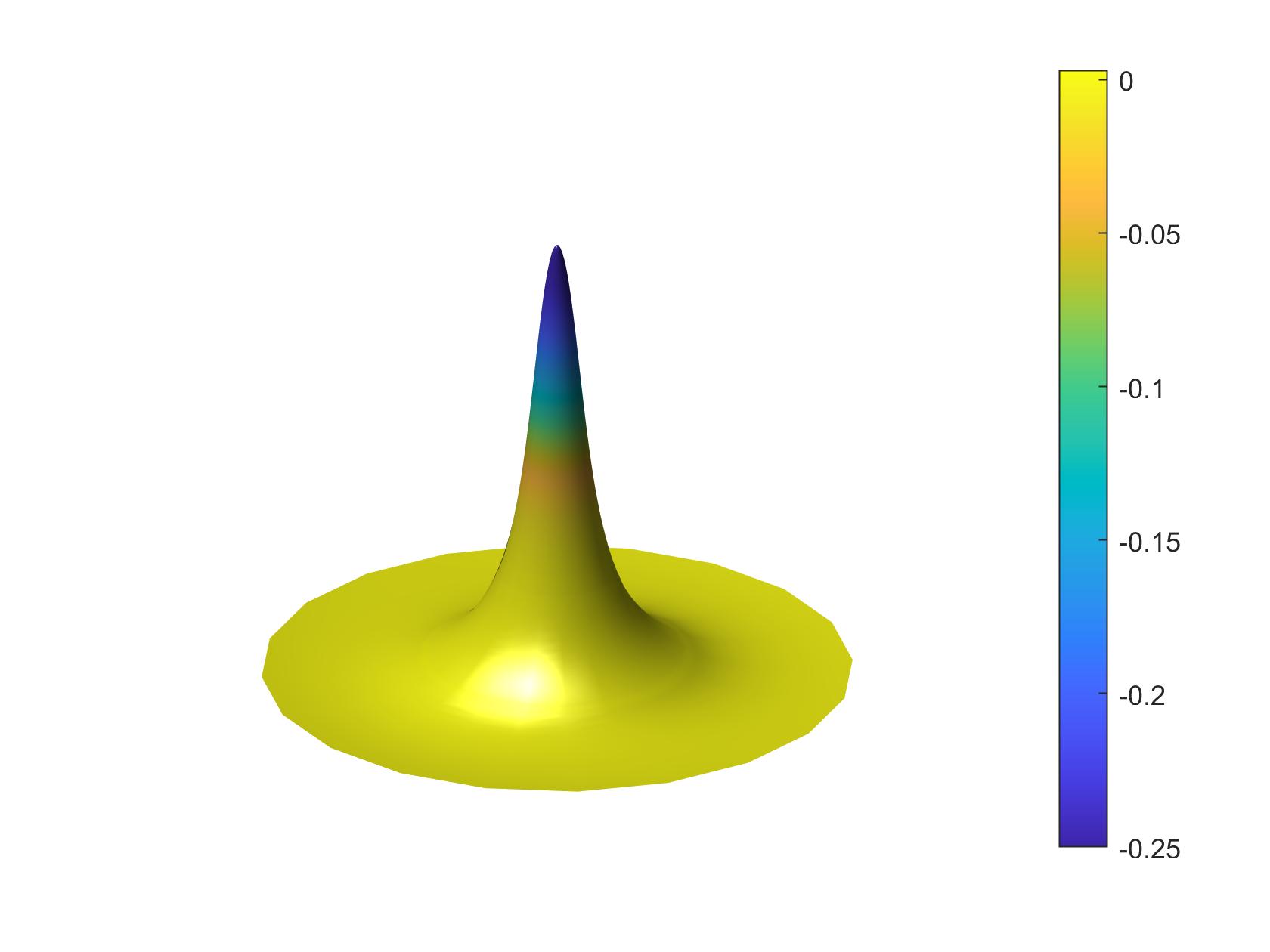}\,
  \,
  \includegraphics[width=5.5cm,height=4.5cm,angle=0]{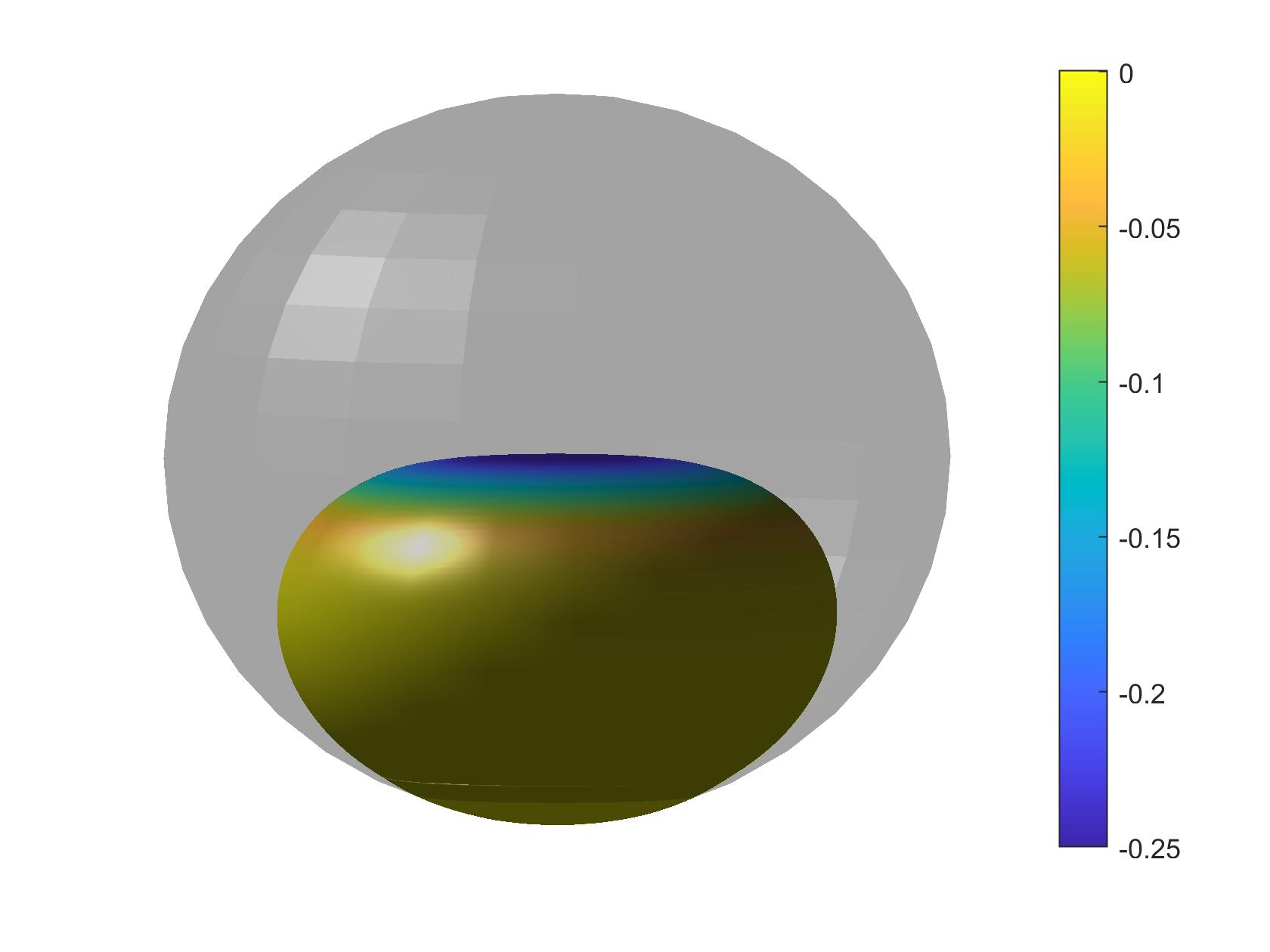}
  \end{center}
 {{\em  {\bf Figure 5.} Regular hyperbolic embeddings of $\Sigma_N$ for $N=2$ in the upper half space and the Poincar\'e ball model, {obtained numerically and colored by the Gaussian Curvature.} }}
\vskip5pt
\begin{center}
\includegraphics[width=5.5cm,height=5.5cm,angle=0]{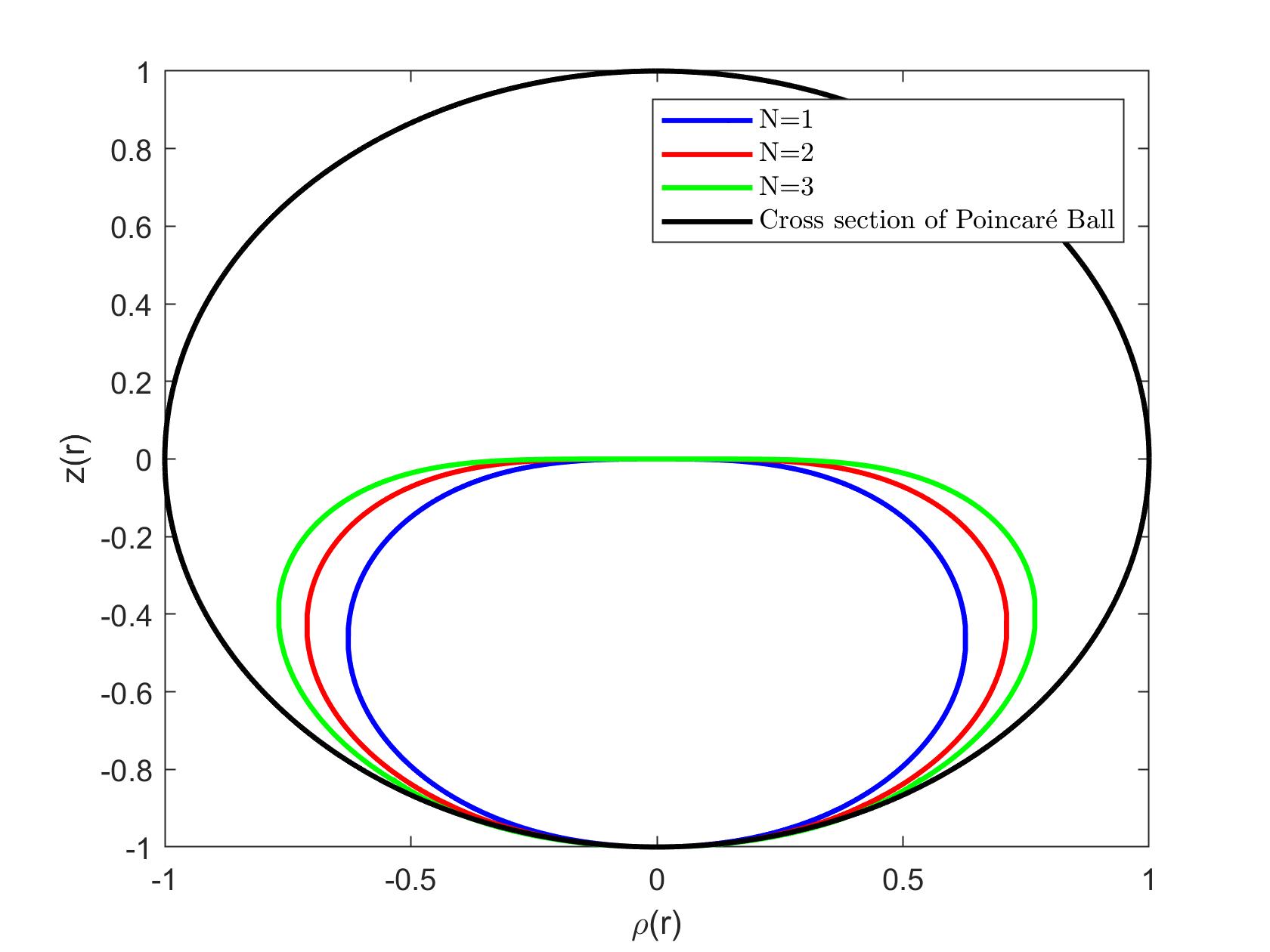}
\end{center}
{{\em  {\bf Figure 6.} Plane sections of the $N=1, 2, 3$ vortex surfaces in the Poincar\'e ball.}}
\vskip 5pt
\item Computing the mean curvature of the hyperbolic embedding (\ref{z_emb}) using the formula
from \cite{DT} yields
\be
\label{mean_first}
H=\frac{-\rho zz'\rho''+\rho z\rho'z''+2((\rho')^2+(z')^2)(\rho\rho'+zz'/2)}{2L\rho((\rho')^2+(z')^2)^{3/2}}.
\ee
 Figure 7 shows the plots of $K$ and $H$ as functions of $r$. The mean curvature $H$ interpolates
between $0$ at $r=0$, which is the mean curvature of the oscullating $\HH^2$, and $-1/2$ which is the mean curvature of the horosphere at $r=\infty$.
\begin{center}
{\includegraphics[width=10cm,height=8cm,angle=0]{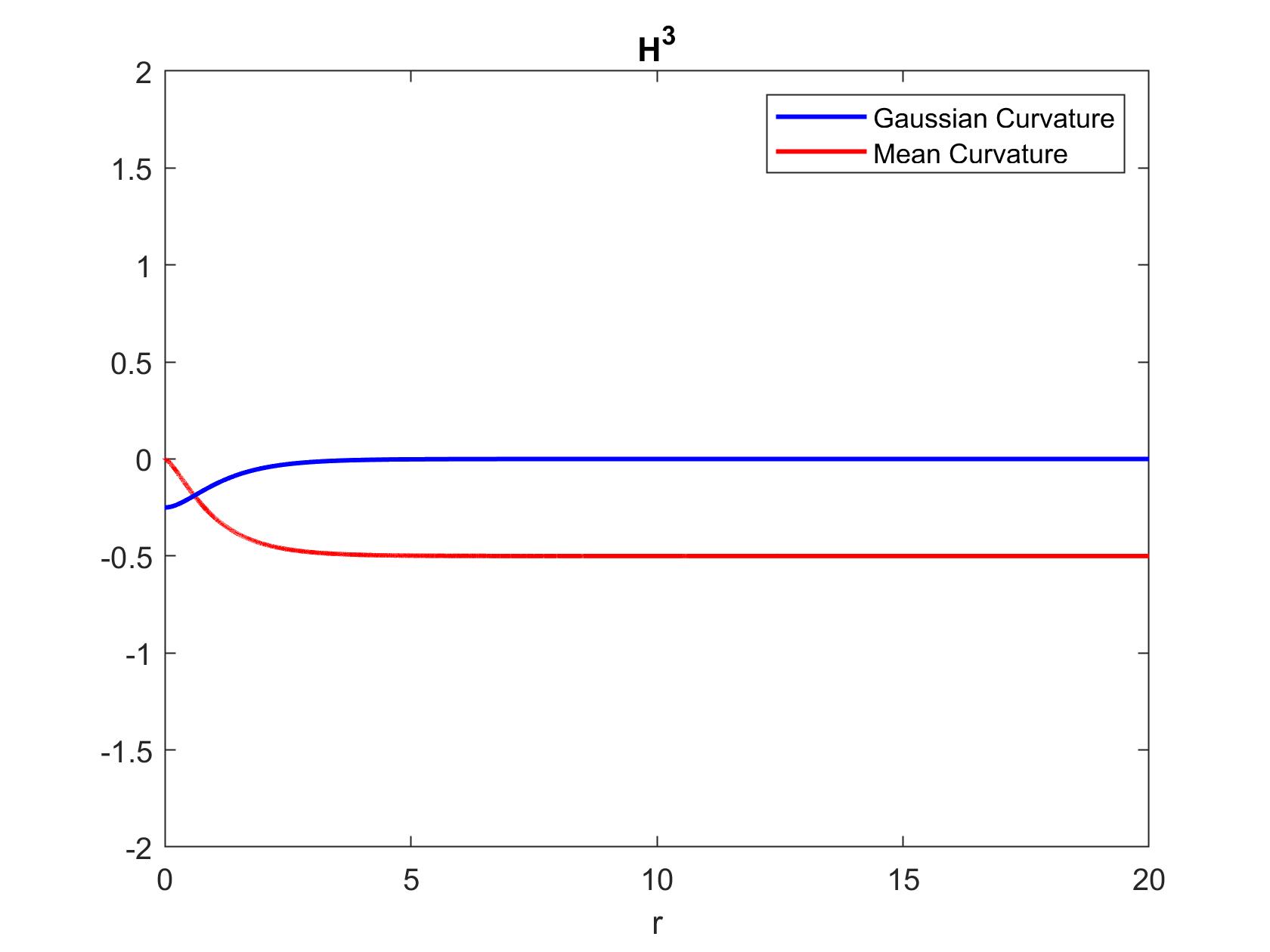}}
\end{center}
{{\em  {\bf Figure 7.} Mean and Gaussian curvatures of the hyperbolic  embeddings for $N=2$.}}
\vskip 5pt
\item Proposition \ref{prop_euclidean_SG} and Theorem \ref{main_theorem} construct a conically singular embedding in $\R^3$, or a regular embedding in $\HH^3$ where both embeddings preserve the intrinsic $U(1)$ isometry.
An alternative would be to seek embeddings/immersions of the $U(1)$ invariant
vortex surfaces, where the embedding does not admit $U(1)$ as a symmetry, i.e. the second--fundamental form is not invariant under $\psi\rightarrow\psi+c$.
Examples of such embeddings are the   
Smyth surfaces \cite{smyth} in $\R^3$ (surfaces of revolution embedded as CMC but only with discrete rotational symmetry), or their counterparts
in $\R^{2, 1}$ also described \cite{brander} by the elliptic Sinh--Gordon equation.
\end{itemize}
\section{Gauss-Bonnet theorem with conical singularities}
Finally we shall re--confirm the vortex number by computing the first Chern class of the bundle ${\mathcal L}$, and relate it to the Gauss--Bonnet
theorem on $\Sigma_N$. Recall that the gauge field $F$ of the Sinh--Gordon vortex, and the Gaussian curvature of the corresponding embedding are given by
\[
F=-\frac{1}{2}\Delta_0 u, \quad K=\frac{1}{4}(e^u-1)
.\]
Remembering that the range of $\theta$ is $[0, (N+2)\pi]$ and integrating the radial Laplacian yields
\begin{eqnarray*}
\frac{1}{2\pi}\int_{\Sigma_N}F&=&-\frac{1}{4\pi}\int_{\Sigma_N}(\Delta_0 u)\; rdrd\theta\\
&=&-\frac{N+2}{4}\Big(\lim_{r\rightarrow\infty} (ru_r)-\lim_{r\rightarrow 0} (ru_r)\Big)\\
&=&N
\end{eqnarray*}
where we used the asymptotic form (\ref{assymp_h}), and (\ref{sssigma}) of $u$ near $r=0$ and the exponential
decay of  the modified Bessel function $K_0(r)$ at $\infty$.
On the other hand the integral of the Gaussian curvature can be related to this Chern class 
\be
\label{GB_N1}
\frac{1}{2\pi}\int_{\Sigma_N}K\mbox{vol}_{\Sigma_N}=-\frac{1}{4\pi}
\int_{\Sigma_N} F=-\frac{N}{2}.
\ee
This is in agreement with the Gauss--Bonnet theorem for surfaces with conical deficit/excess angles \cite{GB_troyanov}. Indeed, if
$g$ is of the form $|z|^{2\alpha_0}|dz|^2$ near $z=0$ and $|w|^{2\alpha_\infty}|dw|^2$ near $w=1/z=0$ then
\[
\frac{1}{2\pi}\int_{\Sigma_N} K \mbox{vol}_{\Sigma_N}=2+\alpha_0+\alpha_\infty.
\] 
In our case $\alpha_0=0$. Setting $\hat{R}=R^{-1}$ in the asymptotic expression for $g$ yields
\[
g\sim \hat{R}^{-N-4}(d\hat{R}^2+\hat{R}^2d\psi^2)
\]
near $\hat{R}=0$. Therefore $\alpha_\infty=-2-N/2$ in agreement with (\ref{GB_N1}).

\section{Tzitzeica vortex} 
There is another possible choice of the conformal factor
in the Taubes equation (\ref{taubes}) which leads to an integrable PDE. Setting
$\Omega=\exp{(-2u/3)}$ yields the
elliptic Tzitzeica equation
\[
\Delta_0 u+e^{-2u/3}-e^{u/3}=0.
\]
The radial solutions $u=u(r)$ of this equation are characterised by Painlev\'e III, this time with parameters $(1, 0, 0, -1)$ (See Appendix).
The asymptotic connection formulae for this
equation have been obtained in \cite{Kit0}. There exists a 
one--parameter family of solutions singular only at the origin.
Adapting the results of \cite{Kit0} to our case  we find
\begin{eqnarray}
\label{assymp_tzi}
u(r)&\sim& \Big(\frac{9p}{\pi}-6\Big)\log{r}+\beta-\frac{e^{-2\beta/3}}{36}\left(1-\frac{p}{\pi}\right)^{-2}r^{6(1-p/\pi)}
\quad \mbox{for}\quad  r\rightarrow 0\\
&\sim& \frac{6\sqrt{3}}{\pi}\Big(\cos{p}+\frac{1}{2}\Big)  K_0(r)\quad \mbox{for}\quad r\rightarrow \infty\nonumber,
\end{eqnarray}
where $0<p<\pi$ parametrises the solutions and
\[
\beta=
3\ln{\Big(3^{-3p/\pi}\frac{9p^2}{2\pi^2}\frac{\Gamma(1-\frac{p}{2\pi})
\Gamma(1-\frac{p}{\pi})}{\Gamma(1+\frac{p}{2\pi})
\Gamma(1+\frac{p}{\pi})}\Big)}-\Big(\frac{9p}{2\pi}-3\Big)\ln{12}.
\] 
Setting
\be
\label{RR2}
r=R^{\frac{3+2N}{3}},\quad\mbox{and}\quad p =\pi \frac{2N+2}{2N+3}
\ee
yields
\[
u=2N\ln{R}+\beta+\dots
\]
which is an $N$--vortex with the strength (compare (\ref{bess})) given by 
\[
\frac{3\sqrt{3}}{\pi}\Big|1-2\cos{\Big(\frac{\pi}{2N+3}\Big)}\Big|.
\]
Near $R=0$ the metric $g=\Omega|dz|^2$ becomes
\[g\sim dR^2+\Big(\frac{3}{3+2N}\Big)^2 R^2d\theta^2.
\]
The constructions of the Euclidean and hyperbolic embeddings  of $g$ proceed along the lines we discussed in the Sinh--Gordon
case, so here we just summarise the results.
\subsection{Euclidean embedding}
The isometric embedding of the resulting surface in $\R^3$ can be
     constructed following the steps in the Proof of Proposition \ref{prop_euclidean_SG}.
     \begin{prop}
     	\label{prop_euclidean_TZ}
       There exists an isometric embedding $\iota: \Sigma\rightarrow \R^3$ of
       the Tzitzeica $N$--vortex as a surface of revolution which is asymptotically
       flat, and regular everywhere  away from the conical singularity {with deficit angle \(4\pi N/(2N+3)\)}.
     \end{prop}
     \noindent
     {\bf Proof.}
     We shall construct this embedding explicitly, assuming that the $U(1)$ symmetry is preserved. We therefore have
     \[
     d\rho^2+\rho^2 d\theta^2+dz^2=[(\rho')^2+(z')^2]dr^2+\rho^2d\theta^2 =e^{-2u/3}(dr^2+r^2d\theta^2)
     \]
so that
     \[
       \rho=re^{-u/3}, \quad z'=\sqrt{\frac{2ru'}{3} e^{-2u/3}\Big(1-\frac{1}{6}ru'\Big)}
     .\]
     For this to exist we need $u'>0$ and $ru'<6$. The expansion (\ref{assymp_tzi}) implies  that $ru'=\frac{6N}{2N+3}$ at
     $r=0$ and $0$ as $r\rightarrow \infty$. Looking for the critical points of $ru'$ we find
     \[
       (ru')'=r(e^{u/3}-e^{-2u/3})
     \]
     which vanishes at $r=0$, or when $u=0$. It is obvious from the boundary conditions at $0$ that $u\neq 0$ as long as $u'>0$ for all $r$, which we need to prove anyway. Note that
     if $u'=0$ and $u>0$ at some point $r_0$ then $u$ has a maximum, and therefore $u''<0$ which contradicts the Tzitzeica equation. The same contradiction is reached if $u'=0$ and $u''=0$.
   If $u'=0$ and $u<0$ then there must also exist a minimum (at some other value of $r$) so that $u$ can reach to $0$ as $r\rightarrow\infty$. The LHS of the Tzitzeica equation is positive at this minimum, but the RHS is negative. In the Proof of (\ref{lemmavh}) (which carries through to the Tzitzeica case) we have also dealt with $u''=0$ at the critical point. Hence $u'$ must be positive for all $r$, which along with the boundary conditions implies $u<0$ for all $r$. It follows that $ru'$ is strictly decreasing, and hence $ru'\leq\frac{6N}{2N+3}<6$ for all $r$. Therefore a global embedding in $\R^3$ exists away from the conical singularity. 
   \koniec
\subsection{Hyperbolic embedding}
The metric $g=e^{-2u/3}|dz|^2$ is
regular at $0$ as long as
$\theta$ is periodic with the period $2 \pi(3+2N)/3$. Taking 
\[
\psi=\frac{3\theta}{3+2N}, \quad r=R^{(2N+3)/3}
\]
makes $\psi$ periodic with period $2\pi$ and the metric becomes flat near 0.
\begin{theo}
	\label{main_theorem_tzitzeica}
	Let $(\HH^3, G)$ be a hyperbolic space with the metric
	$G$ with Ricci scalar $-6/L^2$.
	For any $L^2\in(0,3]$ , there exists a regular isometric embedding $\iota:{\Sigma_N}\rightarrow\HH^3$ of the Tzitzeica N--vortex which preserves the $U(1)$ symmetry of $g_N$, and is unique up to a hyperbolic isometry.
\end{theo}
\noindent
{\bf Proof.}
We follow the proof of Theorem \ref{main_theorem}, and work in the half--space model where the surface of revolution will be constructed with respect to the hyperbolic geodesics
chosen to be the $z$--semiaxis.
Replacing $N$ by $4N/3$ in SG formulae we have
\[
  \iota^*(G)=e^{-2u/3} \Big(dr^2+\Big(\frac{2N+3}{3}\Big)^2 r^2d\psi^2\Big)
\]
which yields the embedding formulae
\begin{eqnarray}
\label{z_emb_t}
  z_\pm&=&z_0\exp{\Big(\int_0^r\frac{2M^2t(tu'-3)/3\pm\sqrt{4M^2e^{2u/3}P}}{2M^2t^2+2L^2e^{2u/3}}dt \Big)},\nonumber\\
\rho_\pm&=&\frac{M}{L}rz_{\pm}e^{-u/3},
\end{eqnarray}
where
\[
  P(r)\equiv {r^2e^{-2u/3}-L^2\left(1-\dfrac{1}{M}-\dfrac{u'r}{3}\right)\left(1+\dfrac{1}{M}-\dfrac{u'r}{3}\right)}, \quad M=\frac{2N+3}{3}\]
with the asymptotic behaviour
\[
  \lim_{r\rightarrow 0} P=0, \quad  \lim_{r\rightarrow \infty} P=\infty
\]
and
\[
  P'(r)=2re^{-2u/3}\left(1-\dfrac{u'r}{3}\right)\left(1+\dfrac{L^2}{3}(e^u-1)\right)
.\]
For the embedding to exist, {we need $P\geq0$}. From the Proof of Proposition \ref{prop_euclidean_TZ}, we have $ru'<\frac{6N}{2N+3}<3$. Moreover, taking the limit $r\to0$ and noting that $e^u\in(0,1)$ we deduce  that $1+\dfrac{L^2}{3}(e^u-1)>0$ iff {$L^2\leq3$}. Hence for this range of $L$, $P'>0$ for all $r$, which implies that $P\geq0$ for all $r$. It follows that there exist two hyperbolic embeddings of the Tzitzeica \(N\)--vortex, and they are related by a hyperbolic isometry as shown in Proposition \ref{prop_2in1}.
\koniec
\subsection{Gauss--Bonnet Theorem with conical singularities}
The Gaussian curvature  of the Tzitzeica $N$--vortex surface is given by 
\[K=\frac{1}{3}(e^u-1).\]
Remembering that $\theta\in[0, 2\pi(2N+3)/3]$ and using the asymptotic behaviour of $u$ given in equation (\ref{assymp_tzi}) we obtain, after integrating the radial Laplacian
\[\frac{1}{2\pi}\int_{\Sigma_N}K\mbox{vol}_{\Sigma_N}=-\frac{2N}{3}.\]
In this case, the metric of the rectified surface \(\Sigma_N\)  is \(dR^2+R^2d\psi^2\), after substituting \(r=R^{2N/3+1}\) and \(\psi=\frac{3\theta}{2N+3}\). Hence \(\alpha_0=0\) and setting \(\hat{R}=R^{-1}\) yields \(\alpha_{\infty}=-2-2N/3\). This is equivalent to the Sinh-Gordon case by interchanging \(N\leftrightarrow4N/3\). Hence 
\[\frac{1}{2\pi}\int_{\Sigma_N}K\mbox{vol}_{\Sigma_N}=2+\alpha_0+\alpha_{\infty}=-\frac{2N}{3}\]
which again agrees with \cite{GB_troyanov}.
\section{Conclusions}
By identifying the norm of the Higgs field in the Abelian Higgs model with a power of a
conformal factor of the underlying background surface we constructed, for each natural numuber 
$N$, regular $N$-vortices corresponding to radial solutions of the Sinh–Gordon and Tzitzeica
equation (\ref{SG_intro}). These are the only two choices leading to integrable PDEs, as it is known \cite{D5, D6} that (up to an overall analytic continuation) the only integrable but not completely solvable PDEs of the 
 form $\Delta_0(u)=G(u)$  are (\ref{SG_intro}). The resulting surfaces $(\Sigma_N, g_N)$ are of separate
interest in the theory isometric embeddings: while they cannot be embedded in $\R^3$
as regular surfaces, global embeddings in $\HH^3$
exist and the embedded surfaces asymptote the CMC
embeddings of horospheres and their covers. If the metric on $\Sigma_N$ is fixed by a radially symmetric solution to the SG or Tzitzeica equation, then the corresponding vortex corresponds t
a point on the moduli space of all $N$-vortices of $\Sigma_N$. While this moduli space exists and has
real dimension $2N$ by the Taubes theorem \cite{taubes_paper}, its detailed construction and the analysis of
the corresponding moduli space metric is an interesting open problem.

\section*{Appendix}
\appendix
\renewcommand{\theequation}{A.\arabic{equation}}
\setcounter{equation}{0}
Painlev\'e III is a family of second order ODEs for $w=w(\zeta)$
parametrised by four constants $(\alpha, \beta, \gamma, \delta)$
\be
\label{piii}
\frac{d^2 w}{d \zeta^2}=\frac{1}{w}\Big(\frac{d w}{d \zeta}\Big)^2  -
\frac{1}{\zeta}\frac{d w}{d \zeta} +\frac{\alpha w^2+\beta}{\zeta}+\gamma w^3 
+\frac{\delta}{w}.
\ee
\begin{itemize}
\item Setting $u=4\ln{w(\zeta)}$ and $r=2\zeta$ in the radial Sinh--Gordon equation
  \[
    u''+\frac{1}{r}u=e^{u/2}-e^{-u/2}
  \]
yields (\ref{piii}) with parameters
  $(0, 0, 1, -1)$
  
  \item Setting $u=3\ln{w(\zeta)}, r=\frac{3\sqrt{3}}{2} \zeta^{2/3}$ in the radial Tzitzeica equation
    \[
      u''+\frac{1}{r}u=e^{u/3}-e^{-2u/3}
      \]
yields (\ref{piii}), this time with parameters
$(1, 0, 0, -1)$. 
\end{itemize}

\end{document}